\documentclass[twocolumn,a4paper]{revtex4}
\usepackage[T1]{fontenc}     
\usepackage{hyperref}
\usepackage{graphicx}
\usepackage{multirow}
\usepackage{color}
\usepackage{amsmath,amsfonts,amssymb}
\let\oldAA\AA
\renewcommand{\AA}{\text{\normalfont\oldAA}}

\newcommand{\tase}{TaSe$_3$}
\usepackage{xr-hyper}
\usepackage{hyperref}
\begin{document}

\title{Strain control of the competition between metallic and semiconducting states in single-layers of TaSe$_3$}

\author{Jos\'e \'Angel Silva-Guill\'{e}n}
\affiliation{School of Physics and Technology, Wuhan University, Wuhan 430072, China}
\email{josilgui@gmail.com}

\author{Enric Canadell}
\affiliation{Institut de Ci\`encia de Materials de Barcelona (ICMAB-CSIC), Campus Bellaterra, 08193 Bellaterra, Spain}

\begin{abstract}
TaSe$_3$ is a metallic layered material whose structure is built from TaSe$_3$ trigonal prismatic chains. 
In this work we report a first-principles density functional theory study of TaSe$_3$ single-layers and we find that, despite the existence of non negligible Se$\cdots$Se interlayer interactions, TaSe$_3$ single layers are found to be metallic.
However, an interesting competition between metallic and semiconducting states is found under the effect of strain. 
The single-layers keep the metallic behaviour under biaxial strain although the nature of the hole carriers changes. 
In contrast, uniaxial strain along the chains direction induces the stabilization of a semiconducting state. 
Potential electronic instabilities due to Fermi surface nesting are found for single-layers under either biaxial strain or uniaxial strain along the long (inter-chain) axis of the layers. 
Bilayers and trilayers have also been considered. 
The structural and electronic features behind these unexpected observations are analyzed.
\end{abstract}


\maketitle

\section{Introduction}\label{sec:intro}

Layered group IV transition metal trichalcogenides (TMT) MX$_3$ (M= Ti, Zr, Hf; X= S or Se) have recently attracted much interest because in the form of single-layer or few-layers thick they may provide a convenient platform for applications in advanced devices~\cite{dai2016,island2017,jin2015,dai2015}. 
These solids crystallise in the ZrSe$_3$-type structure~\cite{furuseth1975,furuseth1991} with layers built from trigonal prismatic chains of chalcogen atoms containing a transition metal atom in the center of every trigonal prism. 
With the exception of the tellurium compounds all solids of this family are semiconducting. 
These chalcogenides are thus structurally anisotropic~\cite{furuseth1975,furuseth1991} and the bulk indirect band gap may change to direct (TiS$_3$) or keep its indirect nature (TiSe$_3$) in the single-layers~\cite{dai2015,jin2015,li2015,kang2016}. 
The bottom/top part of their conduction/valence bands have opposite curvature~\cite{dai2015,li2015,silva2017} and this property opens the possibility to reverse the anisotropy of the electrical and optical properties by switching from $n$- to $p$-doping. 
Recently, we have proposed that the anisotropy of the conduction band can also be reversed in TiS$_3$ by moderate compressive strain~\cite{silva2018}. 
These features may be used to build nanostructures with switchable plasmon channeling. 
The variation of the band gap in single-layers when applying strain or changing the nature of either the transition metal or chalcogenide atoms, the tuning of the mechanical and optoelectronic properties, and the role of vacancies in flakes or nanoribbons of this family  have been extensively studied~\cite{dai2016,island2017,dai2015,jin2015,li2015,kang2016,silva2017,silva2018,iyikanat2015,island2014,island2015,lipatov2015,kang2015}.

\begin{figure}[!b]
    \centering
    \includegraphics{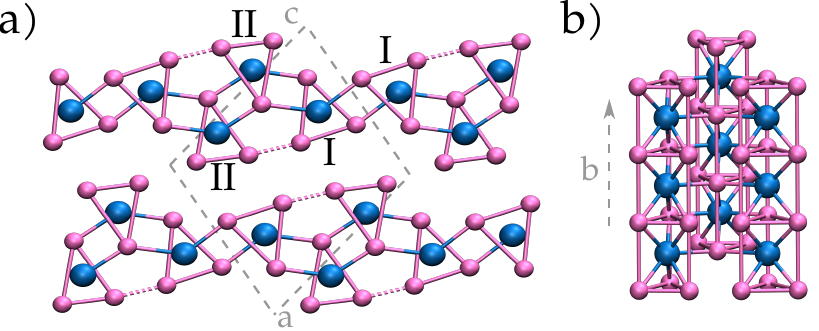}
    \caption{Crystal structure of \tase~at room temperature. The labels I and II refer to the two different types of chains discussed in the text. The Ta and Se atoms are shown as blue and purple spheres, respectively. }
    \label{fig:structure}
\end{figure}

Although the crystal structure is also built from layers of trigonal prismatic chains, the group V TMT are very different in terms of both structure and physical properties. 
Some of the members of this family, like NbSe$_3$ and monoclinic TaS$_3$, rank among the more intensely studied low-dimensional metals because of the intriguing physics associated with their charge density wave (CDW) instabilities~\cite{Monceau2012}. 
The electronic structure of these  materials is strongly dependent upon the structural features of the different trigonal prismatic chains found in their crystal geometry. 
In contrast with the group IV TMT, which exhibit only one type of chain, the presently known group V TMT can exhibit one (NbS$_3$ polymorph I)~\cite{Rijnsdorp1978} two (TaSe$_3$)~\cite{Bjerkelund1966}, three (NbSe$_3$)~\cite{Meerschaut1975,Hodeau1978} or even four (NbS$_3$ polymorph II)~\cite{Zupanic2018} cristallographically different trigonal prismatic chains. 
Whereas the chalcogen atoms of the triangular units are well separated from each other in some of these chains so that they must be formally considered as X$^{2-}$, in other chains two of the chalcogen atoms form a single X-X bond so that the three chalcogen atoms must then be considered as (X$_2$)$^{2-}$+X$^{2-}$. This subtle structural feature imposes the occupation of the transition metal based bands and completely determines the physical properties of the system~\cite{Monceau2012}. 
Despite the extremely interesting physics of these solids in the bulk, to the best of our knowledge they have not yet been carefully studied as few-layers flakes. 
Here, we would like to call attention towards one of these solids, TaSe$_3$ which according to our calculations could exhibit an appealing behaviour as a single-layer.   

TaSe$_3$ crystallizes in the monoclinic system~\cite{Bjerkelund1966} and exhibits TaSe$_3$ layers with a repeat unit of four trigonal prismatic chains (see Fig.~\ref{fig:structure}). 
Two of the chains, noted I and II, are crystallographically independent. 
The Se-Se distance parallel to the layer direction in chain II (2.575\AA) is compatible with a single bond, whereas that in chain I is too long (2.896 \AA). 
Consequently, the system can be formulated as 2 $\times$ [Ta$_{I}$(Se$^{2-}$)$_3$ + Ta$_{II}$(Se$^{2-}$)(Se$_2^{2-}$)]. 
In other words, all Ta atoms are formally $d^0$ and thus there are no electrons to fill the low-lying transition metal-based bands of the four \tase~chains and a semiconducting behaviour could be expected. However, because of the short inter-chain Se$\cdots$Se contacts (one of these contacts, connecting chains I and II and noted as a broken line in Fig.~\ref{fig:structure}a, is unusually short, 2.653 \AA) which raise up the Se-based valence band, the top of the valence band and the bottom of the conduction band overlap~\cite{Perucchi2004,Canadell1990} and TaSe$_3$ exhibits non-activated conductivity at room temperature. 
As a matter of fact, TaSe$_3$ keeps the metallic conductivity until very low temperatures and at around 2 K enters into a superconducting state~\cite{Sambongi1977,Haen1978,Fleming1978,Yamamoto1978,Yamaya1979,Nagata1989,Tajima1984} whose nature is not yet well understood.  
Interestingly, TaSe$_3$ has also been grown as microwires~\cite{Stolyarov2016} and with unusual shapes like a M\"obius strip~\cite{Tanda2002,Matsuura2003}.  

Very recently there has been a revival of interest in TaSe$_3$ and appealing properties like high breakdown current density~\cite{Stolyarov2016,Empante2019} and low-frequency electronic noise~\cite{Liu2017} in microwires have been reported suggesting potential applications in downscaled electronics. In contrast with other solids of this family no CDW has been observed in single crystals. 
However, the occurrence of a CDW state has been claimed to occur in mesowires at 65 K~\cite{Yang2019}. 
It has also been proposed that bulk TaSe$_3$ may be an appropriate system where to study the competition between superconducting and topological phases~\cite{Nie2018}. 
It has been reported very recently that single-layer TaSe$_3$ nanoribbons on SiO$_2$/Si substrates can be obtained through a mechanical exfoliation procedure~\cite{Kim2019}. 
Thus, as part of this resurgence of interest in TaSe$_3$, we would like to call attention here towards the remarkable strain-induced behaviour of TaSe$_3$ as a single-layer.        

\begin{figure}[t!]
    \centering
    \includegraphics{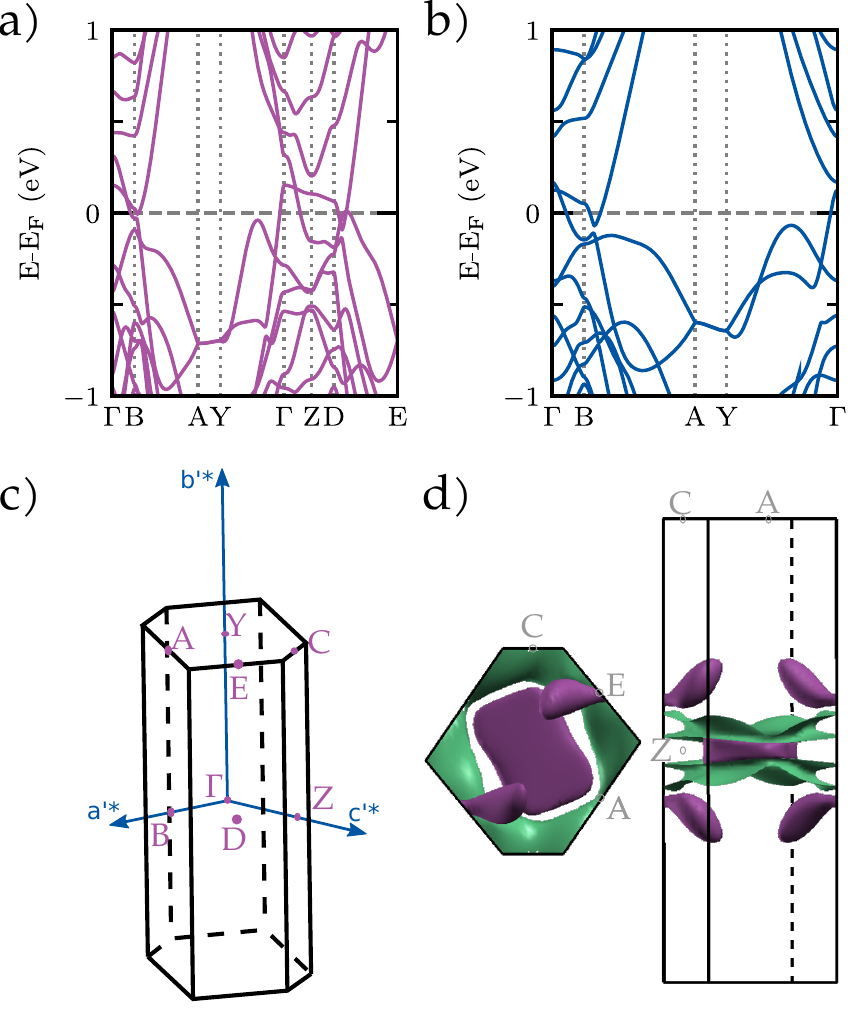}
    \caption{Band structure of bulk (a) and monolayer (b)~\tase. Note that a non-conventional set of axes ($a$'= $a$+$c$, $b$'= $b$ and $c$'= $c$), where $a$' and $b$' run along the two repeat vectors of the~\tase~layer has been used. (c) Brillouin zone, and (d) top and side views of the calculated Fermi surface for bulk~\tase. Color code: $\mathrm{violet}=\mathrm{holes}$, $\mathrm{green}=\mathrm{electrons}$.} 
    \label{fig:mono-bulk-comp}
\end{figure}

\section{Results}\label{sec:results}

\begin{figure*}
    \centering
    \includegraphics{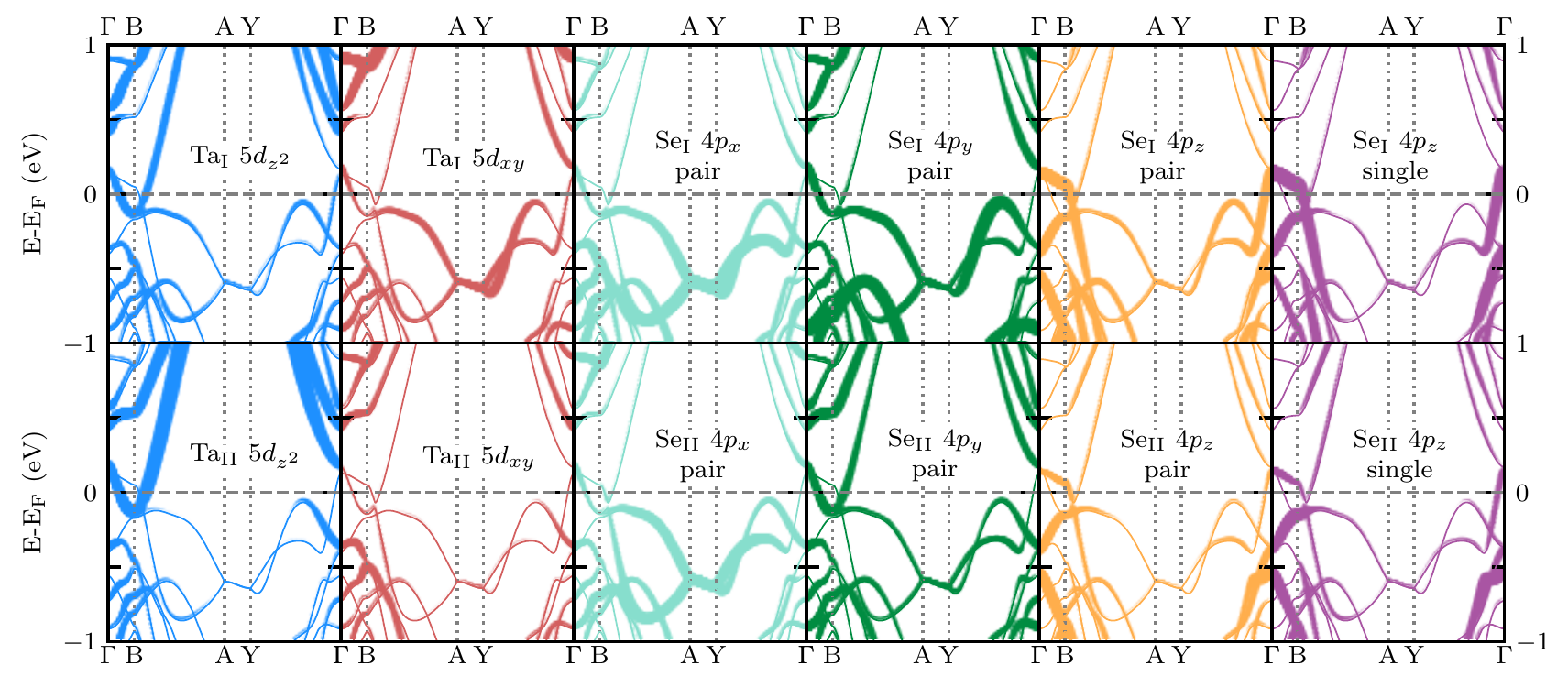}
    \caption{Fatband analysis of the TaSe$_3$ single-layer band structure. The labels I and II refer to the two different trigonal prismatic chains (see Fig.~\ref{fig:structure}). The label "pair" refers to the two Se atoms forming the bond in chain II and the equivalent ones in chain I. The label "single" refers to the third Se atom of the triangles.}
    \label{fig:fatbands_reduced}
\end{figure*}

As discussed above, because of the internal structure of the trigonal prismatic chains, the Ta atoms of TaSe$_3$ are formally in a $d^0$ configuration so that a semiconducting behaviour could be expected.
However, a metallic (and low temperature superconducting) state is observed. Our calculated band structure for bulk TaSe$_3$ (Fig.~\ref{fig:mono-bulk-comp}a) is in agreement with this observation, as well as with previous band structures~\cite{Canadell1990,Nie2018}. 
Note that the TaSe$_3$ layers occur along the ($a$+$c$)- and $b$-directions (see Fig.~\ref{fig:structure}). Consequently, in order to facilitate our discussion we have used a non-conventional set of axes, $a$'= $a$+$c$, $b$'= $b$ and $c$'=$c$, where $a$' and $b$' run along the long and short repeat vectors of the~\tase~layer and $c$' along the inter-layers direction.
As shown in Fig.~\ref{fig:mono-bulk-comp}a, the electron and hole pockets of bulk TaSe$_3$ result from the hybridization of two different bands, a mainly filled quite 1D band originating from the Se orbitals (mostly 4$p_z$) and a mainly empty band originating from Ta $d_{{z^2}}$ orbitals. 
The latter is associated with strong interactions along the chains direction, but it also exhibits a sizeable dispersion along the inter-chains direction. 
The two bands hybridize in some sections of the Brillouin zone (see later for further discussion when considering the single-layers) through the mutual overlap with Se 4$p_y$ orbitals of the same and/or adjacent chains. 
The former band leads to the purple pancake-like around $\Gamma$ of the Fermi surface  (Fig.~\ref{fig:mono-bulk-comp}d) together with some additional closed purple pockets, whereas the latter leads to a pair of green warped double sheets with big holes around the pancake. 
Despite the hybridization, most of the hole pockets are associated with the Se 4$p_z$ orbitals and the electron pockets are mostly associated with the Ta $d_{{z^2}}$ orbitals. 
Small changes in the semi-metallic band overlap only slightly change the size of the pancake and may break the green warped sheets with holes into smaller fragments. 
An interesting observation is that around the Fermi level there are non-negligible inter-layer interactions. 
For instance, for any reasonable value of the Fermi level the pancake remains, so that even if the conductivity is stronger along the $b$-direction, it should be quite isotropic within the ($a$,$c$)-plane. 
This feature suggests that the physical behaviour may change in the absence of Se$\cdots$Se inter-layer interactions and we decided to study the electronic structure of TaSe$_3$ single-layers.  

The optimized structure of a single-layer is essentially the same as for the bulk and the cell parameters are within 3\% those of the bulk experimental crystal structure. 
The calculated band structure is reported in Fig.~\ref{fig:mono-bulk-comp}b. 
We carried out additional calculations including spin-orbit coupling but, except for the appearance of slightly avoided crossings, the band structure around the Fermi level (see Fig. S1 in Supplementary Information (SI))  does not exhibit substantial differences. 
Consequently, from now on all reported band structures will be those calculated without spin-orbit coupling. 
Interestingly, at first glance, we can see that the first important result of Fig.~\ref{fig:mono-bulk-comp}b is that the TaSe$_3$ single-layer is also metallic.
Moreover, looking carefully to the band structure, we can see that the main characteristics of the bulk band structure near the Fermi level are kept. 
There are holes around the $\Gamma$ point, a slightly avoided crossing between two strongly dispersive bands near the B point and the presence of a band with a hump shape slightly below the Fermi level along the $\Gamma \rightarrow$ Y direction. 
This leads us to the conclusion that the main reason for the metallic behaviour of TaSe$_3$ does not originate from inter-layer interactions, even if they are non-negligible, but from the inner structure of the layers.

Since our purpose is the analysis of possible modifications of the electronic structure under strain we need take a deeper look at the origin of the three features of the electronic structure noted above. 
Their evolution will point out whether the possibility to open a band gap at the Fermi level is likely or not and how the nature of the metallic state may evolve under strain. 
A simple and meaningful approach is to carry out a fatband analysis of the bands near the Fermi level. Shown in Fig.~\ref{fig:fatbands_reduced} are the essential contributions needed for our goal. 
None of the other contributions influences the bands near the Fermi level but the full analysis may be found in the SI (Figs. \ref{SI-fig:fatbands_I} and \ref{SI-fig:fatbands_II}). 
Since there are two different trigonal prismatic chains (I and II, see Fig.~\ref{fig:structure}a), we have separated the Ta and Se contributions for each type of chain. 
We recall that there is a Se--Se bond in chain II but not in chain I. 
The orbitals are specified according to a system of axes where the $x$- and $z$-axes run along the long (i.e. $a$+$c$ in the bulk) and short (i.e. $b$ in the bulk) repeat vectors of the layer. 
For the Se atoms we use the labels "pair" and "single" to distinguish the two Se atoms that form the bond in chain II and the equivalent ones in chain I (Se$_{{pair}}$) from the third Se atom of the triangles (Se$_{{single}}$). 

The electron pockets originate from the Ta 5$d_{z^2}$ orbitals. 
This is expected because for a transition metal atom in a trigonal prismatic coordination there are three low-lying $d$ orbitals: $d_{z^2}$, $d_{xy}$ and $d_{x^2-y^2}$. 
Since the Ta atom of one chain lies on the same plane as the Se atoms of the two neighboring chains (Fig.~\ref{fig:structure}b), the 5$d_{xy}$ and 5$d_{x^2-y^2}$ orbitals strongly interact with the 4$p_x$ and 4$p_y$ of the two Se capping atoms and these two $d$ orbitals are pushed to higher energies. 
Consequently, the only low-lying Ta $d$ orbital remaining is the 5$d_{z^2}$. 
However, those of the Ta atoms in chains II clearly dominate because the occurrence of a short Se--Se bond stabilizes the 5$d_{z^2}$ orbitals of this chain~\cite{hoffmann1980,Canadell1990,Canadell1989} by decreasing the hybridization between the 5$d_{z^2}$ and Se--Se bonding orbitals. 
Note that this prevalence is consistent with the electronic structure of other TMT like NbSe$_3$ and TaS$_3$ although the layers are somewhat different~\cite{Canadell1990,Nicholson2017,Valbuena2019}. 
The second band crossing the Fermi level is a strongly raising band along the Y $ \rightarrow \Gamma$ and A $ \rightarrow $ B directions (i.e. along the chains direction). 
This band is dominated by the Se 4$p_z$ orbitals of the three Se atoms of type I chains. 
These orbitals make strongly antibonding interactions along the chain direction when approaching the B and $\Gamma$ points where the successive 4$p_z$ orbitals become in-phase and, thus, make strong  antibonding interactions along the chains.  
Those of chain I are higher in energy for two reasons. First, in chain I there is no short contact and, consequently, no bonding interaction occurs between the Se$_{{pair}}$ 4$p_z$ orbitals so that these orbitals are higher in energy. 
Second, the Se$_{{single}}$ atoms of chain I make one Nb--Se bond with the adjacent chain, whereas those of chain II make two bonds. 
As a result, the interaction of the Se 4$p_z$ with the Ta 5$d_{xz}$ and 5$d_{yz}$ orbitals, which is allowed by symmetry and stabilizes the Se 4$p_z$ orbital, is weaker in chain I, which results in that the Se$_{{single}}~$4$p_z$ band is higher in energy. 
Consequently, the top of the valence band is dominated by the chain I Se orbitals. 
The avoided crossing between the Ta 5$d_{z^2}$ and Se 4$p_z$ bands near the B point is only weakly avoided because these bands are mostly located in different chains.

\begin{figure}
    \centering
    \includegraphics{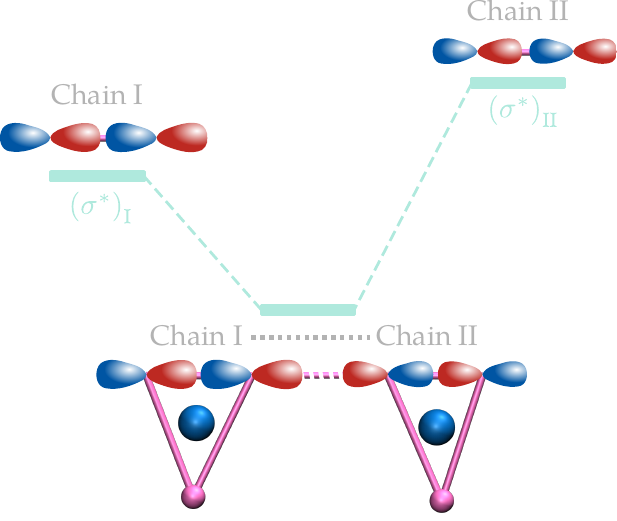}
    \caption{Schematic representation of the component along the direction of the pair Se-Se atoms of the hump-shaped band just below the Fermi level along the Y$\rightarrow \Gamma$ direction.}
    \label{fig:hump_orbital}
\end{figure}

\begin{figure}
    \centering
    \includegraphics{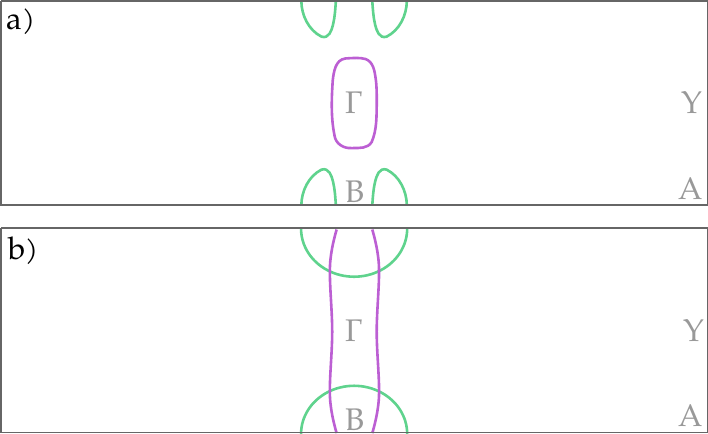}
    \caption{(a) Calculated Fermi surface for a TaSe$_3$ single-layer. Color code: $\mathrm{violet}=\mathrm{holes}$, $\mathrm{green}=\mathrm{electrons}$. (b) Schematic illustration of the  open hole and closed electron Fermi surfaces whose hybridization leads to the Fermi surface in (a).}
    \label{fig:FS_I}
\end{figure}

\begin{figure}
    \centering
    \includegraphics{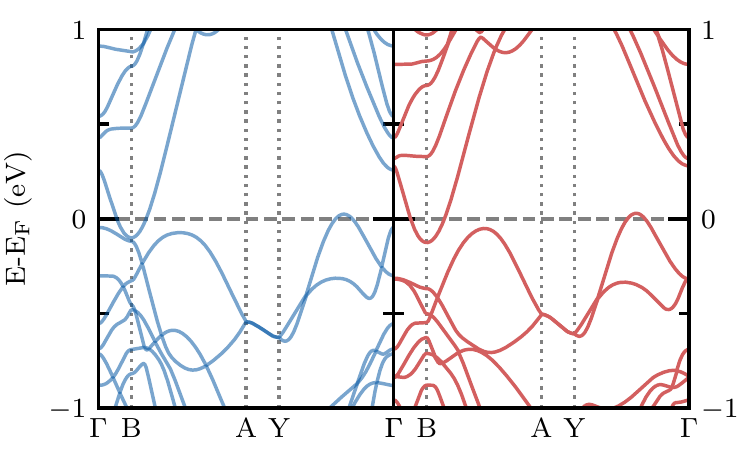}
    \caption{Band structure of monolayer \tase~with a biaxial stress of 2\% (left) and 5\% (right) \tase.}
    \label{fig:biaxial}
\end{figure} 

\begin{figure}
    \centering
    \includegraphics{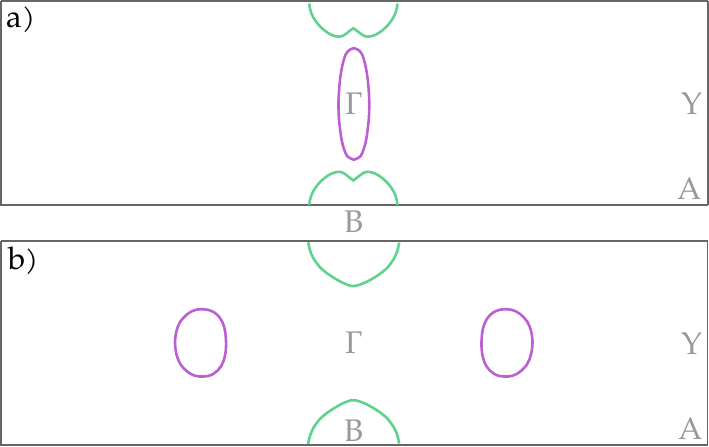}
    \caption{ Calculated Fermi surface for a TaSe$_3$ single-layer under biaxial strain of 1 \% (a) and 2.5 \% (b). Color code: $\mathrm{violet}=\mathrm{holes}$, $\mathrm{green}=\mathrm{electrons}$.}
    \label{fig:FS_II}
\end{figure}

\begin{figure*}
    \centering
    \includegraphics{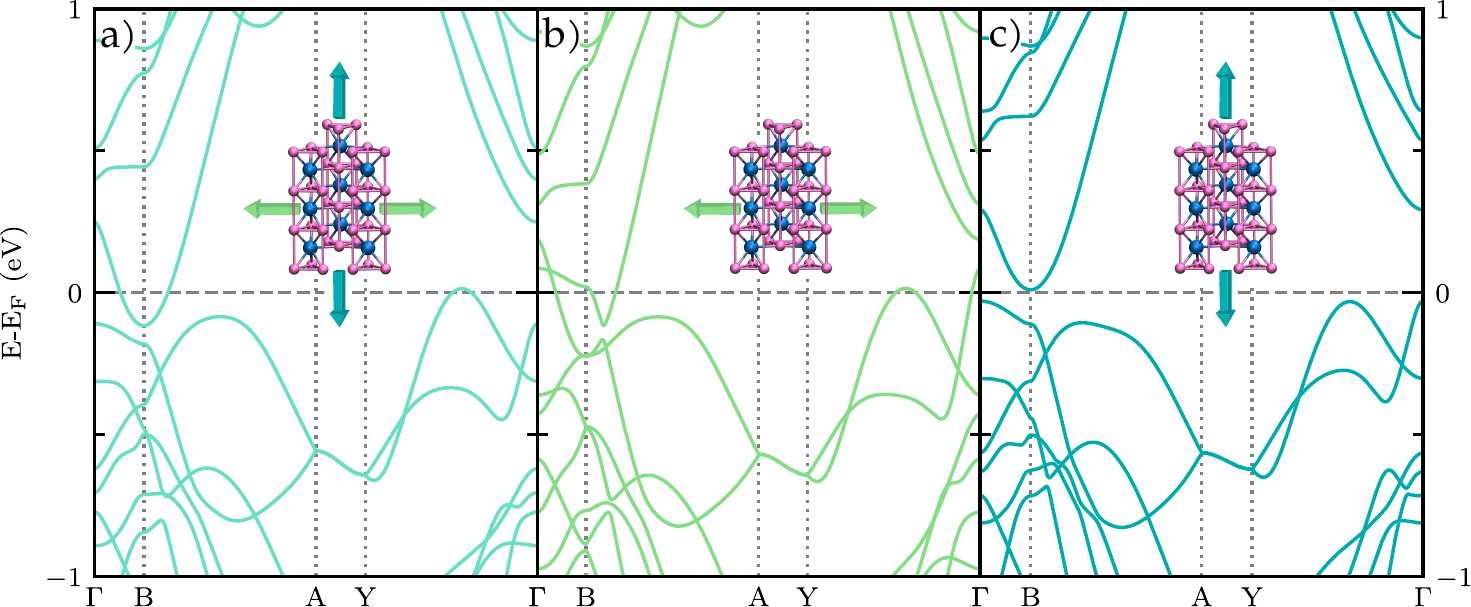}
    \caption{Comparison of the band structure of a \tase~single-layer with different types of applied strain: a) biaxial, b) uniaxial along the interchain $a$'-direction and c) uniaxial along the chain $b$'-direction. In all cases the applied strain was $\epsilon=2.5\%$.}
    \label{fig:all-type-strains}
\end{figure*}

There is a third band that, even if not crossing the Fermi level, must be taken into account when considering the potential modifications induced by strain: the hump-shaped band along the Y $\rightarrow \Gamma$ direction. 
This band is based on the orbitals of the Se$_{{pair}}$ atoms and can be decomposed in two contributions, a component pointing outside the layer (i.e. made of the Se 4$p_y$ orbitals of mostly chain I) and a component along the direction of the Se-Se pairs (i.e. made of the 4$p_x$ orbitals) equally shared by chains I and II. 
Let us look in some detail at this component since it will play an important role when applying strain. 
As mentioned in the introduction, the contact drawn as a broken line in Fig.~\ref{fig:structure}a, which connects chains I and II, is unusually short, 2.653 \AA. In fact, it is intermediate between the short and long Se-Se pairs (2.575 and 2.896 \AA, respectively) so that from the electronic viewpoint the four Se atoms should be considered as a tetrameric unit. The component we are discussing is built from the combination of the four Se 4$p_x$ orbitals of this unit and is schematically drawn in the lower part of Fig.~\ref{fig:hump_orbital}. It is the second highest orbital of the set of four that can be generated from the Se 4$p_x$ orbitals of the tetrameric unit. A simple way to understand the shape of this orbitals is to consider it as resulting from the interaction of two $\sigma$* orbitals of the short and long Se-Se pairs of chains II and I, respectively (see Fig.~\ref{fig:hump_orbital}). Since the Se--Se distance in chain I is longer, the  $\sigma$*$_I$ is lower in energy than $\sigma$*$_{II}$ and thus, they interact in a bonding way between the pairs to give the lower-lying combination. This is the component along the direction of the paired Se-Se atoms of the hump-shaped band just below the Fermi level along the Y$\rightarrow \Gamma$ direction, which will play an important role under strain application. 

The Fermi surface associated with the TaSe$_3$ single-layer in the absence of strain is shown in Fig.~\ref{fig:FS_I}a. Although the slightly avoided crossing near B may be a bit confusing, this Fermi surface can be considered to result from the hybridization (see Fig.~\ref{fig:FS_I}b) of (i) a pseudo-1D hole Fermi surface around $\Gamma$ mostly coming from the chain I Se 4$p_z$ orbitals and (ii) a closed electron Fermi surface centered at the B point mostly due to the chain II Ta 5$d_{z^2}$ orbitals. The result is a rounded rectangular hole pocket around $\Gamma$ and two lentil-like closed electron pockets of half the area of the hole pocket which are centered in the B $\rightarrow$ A line. Note that this Fermi surface is consistent with the bulk one where the Fermi level is slightly lowered, i.e. when the chain I Se 4$p_z$ band is slightly lowered because of decreased Se$\cdots$Se interactions.  

Based on these results we came to the conclusion that going from bulk to the single-layer, even if decreasing the Se$\cdots$Se interactions which raise the top of the valence band, is not enough to change the conductivity regime of TaSe$_3$. 
However, the area of the hole pocket is only 2.6\% of the Brillouin zone (See Fig.~\ref{fig:FS_I}a) and given the topology of the band structure, strain should provide a practical way to alter the situation. 
Thus, we have calculated the evolution of the electronic structure as a function of applied biaxial strain. Shown in Fig.~\ref{fig:biaxial} are the results for 2\% and 5\% strain. 
Surprisingly, we did not observe the opening of a band gap. 
As a matter of fact, even for a biaxial strain of 7.5\%, the metallic character is kept. 
However, it is important to point out that the nature of the metallic state has changed. 
The calculated Fermi surface for a 2.5\% strain is shown in Fig.~\ref{fig:FS_II}b (the band structure and FS for a 5\% strain are shown in Figs. \ref{SI-fig:biaxial_SO} and \ref{SI-fig:FS_biax_5} of SI). 
Whereas the electron pockets are still due to chain II Ta 5$d_{z^2}$ orbitals, the holes do not occur anymore at $\Gamma$ but along the Y $\rightarrow \Gamma$ line. 
In fact, in this case there is just one electron pocket but two identical hole pockets. 
Note that the two hole pockets are well nested by a $q \sim \upsilon  b\mathrm{'}^{*}$ nesting vector (the $\upsilon$ value changes from 0.43 to 0.46 from 2\% to 5\% biaxial strain) so that the single layer could exhibit a metal-to-metal transition associated with a CDW modulation along the chains direction due to the nesting of the hole pockets.  

The effect of strain along the chains direction is that both Ta--Ta and Se--Se distances increase (for 2.5\% strain they both lengthen by 0.089 \AA) so that both the chain II Ta 5$d_{z^2}$--Ta 5$d_{z^2}$ bonding interactions and the chain I Se 4$p_z$--Se 4$p_z$ antibonding interactions decrease. 
Consequently, the corresponding bands separate and, in fact, at 2.5\% they do not overlap anymore. 
However, because of the strain along the inter-chains direction, the shorter inter-chain contact between chains I and II (see Fig.~\ref{fig:structure}a) becomes longer. 
According to our calculations, a modest 2.5\% strain lengthens this contact by $\sim$0.1~\AA , whereas the two other Se--Se contacts change considerably less. 
Since, as analyzed above, the levels at the top of the hump-shaped band along the Y $\rightarrow \Gamma$ direction are locally bonding for this inter-chain contact, strain destabilizes these levels which raise up and cross the Fermi level, leading to the creation of the new holes. 
Because of the raising of the hump-shaped band, which already for a weak biaxial strain of less than $\sim$2\% crosses the Fermi level, there is always a semimetallic overlap. 
Shown in Fig.~\ref{fig:all-type-strains} are the calculated band structures for the cases of biaxial and uniaxial strains of 2.5 \% showing that biaxial or uniaxial strain along the inter-chains direction keep the metallic behaviour, whereas a band gap is induced through uniaxial strain along the chains direction. 
Thus, under biaxial strain a metallic state is retained. 
However, the nature (but not the shape) of the electron pockets is kept, although the nature of the hole pockets changes from Se 4$p_z$ in chain I to Se 4$p_x$ in both chains.

\begin{figure}[t!]
    \centering
    \includegraphics{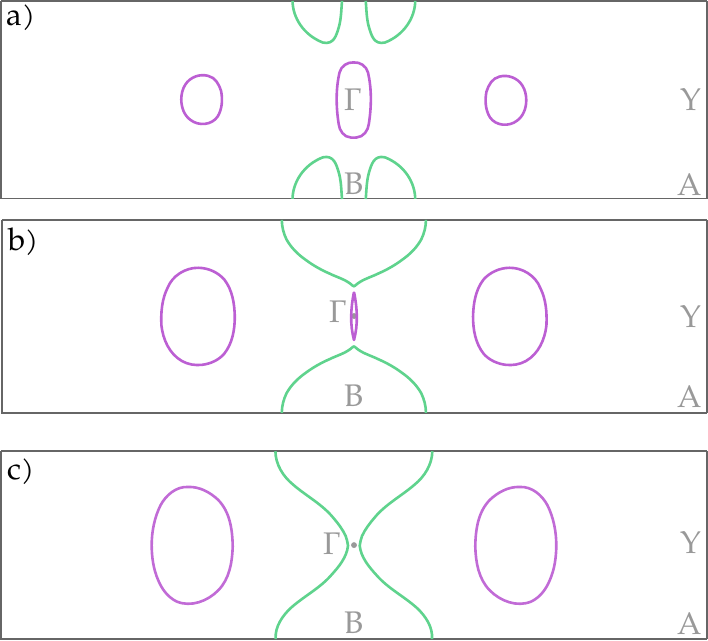}
     \caption{Evolution of the Fermi surface of a \tase~single-layer under uniaxial strain of 2.5 \% (a), 5 \% (b) and 7.5 \% (c) along the long axis ($a$') of the layer. Color code: $\mathrm{violet}=\mathrm{holes}$, $\mathrm{green}=\mathrm{electrons}$.}
    \label{fig:uniax_a}
\end{figure}

It is clear from the previous analysis that a semiconducting state should be attainable by applying uniaxial strain along the chains direction. 
According to the present calculations a modest strain of $\sim$~2.5\% is enough to induce it and it is kept for higher strain (see Fig. \ref{SI-fig:bs-uniaxial-a-5} in SI).
Moreover, the gap of the semiconducting state can be tuned up to 0.18 eV if we increase the tensile strain up to 5\% (see Fig. \ref{SI-fig:bs-uniaxial-a-5} in SI). 
On the other hand, comparison of Figs.~\ref{fig:FS_I}a,~\ref{fig:FS_II}a and~\ref{fig:FS_II}b clearly show the transition between two metallic states under biaxial strain. 
Initially, the chain I Se 4$p_z$ holes become narrower and more elongated while the two chain II Ta 5$d_{z^2}$ electron pockets collapse. 
At the same time new holes based on Se 4$p_x$ orbitals of both chains emerge in a completely different region of the Brillouin zone. 
The transit between the two situations occurs for a 1-2\%  biaxial strain. 
Whereas the effective mass of holes and electrons for the metallic state of single-layers without strain are very similar, those for holes are substantially smaller for the metallic state under biaxial strain of 2-7.5\%. 
We have checked that introduction of spin-coupling effects do not alter the analysis of the strain effect (see Fig. \ref{SI-fig:all-type-strains-so} in SI). 
Still, a third metallic state could be attained by injecting electron carriers in the conduction band of the semiconducting state through electric field gating. 
In that case, the Fermi surface would be simply a closed electron pocket centered at B, whatever the electron doping is. 
The effect of uniaxial strain along the interchain direction may be ascertained from the evolution of the Fermi surface for strains up to 7.5\% shown in Fig.~\ref{fig:uniax_a}. 
Now the hump-shaped band crosses the Fermi level sooner because the intra-chain interactions are practically not altered. 
For weak strains both types of hole pockets discussed above are found and, for strains slightly larger than 5\%, the Fermi surface is like that for 2.5\% biaxial strain but with an area of the hole and electron pockets considerably larger. 
For near 7\% strain the electron pockets become open warped lines very well nested by a $q_I \sim (1/2)a\mathrm{'}^{*}+  \upsilon b\mathrm{'}^{*}$ nesting vector, i.e. implicating a doubling of the periodicity along the inter-chain direction and an incommensurate modulation along the chains direction. 
Again, we note that the hole pockets along the Y $\rightarrow \Gamma$ direction are very well nested by a $q_{II} \sim \upsilon b{'}^{*}$ nesting vector (the $\upsilon$ value changes from 0.43 to 0.46 from 2\% to 7.5\% uniaxial strain). Thus, the metallic state under strain along the long direction of the layer is susceptible to exhibit two different types of CDW anomalies as a function of applied strain although in both cases they are of the metal-to-metal type because they only concern one type of carriers.   

\begin{figure}
    \centering
    \includegraphics{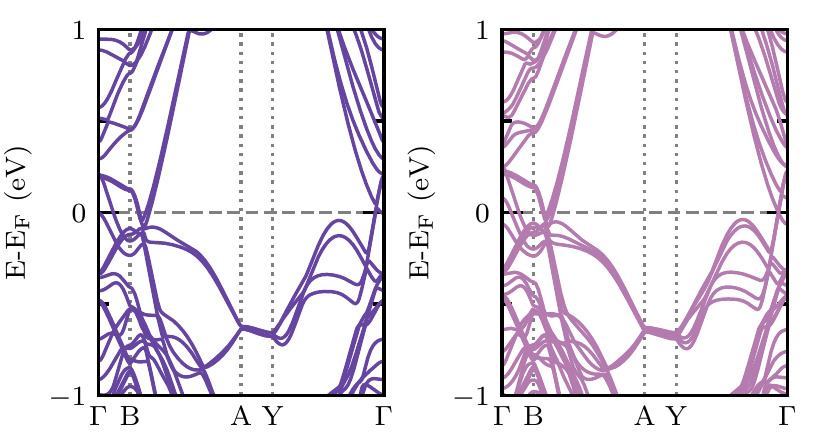}
     \caption{Band structure of bilayer (left) and trilayer (right) \tase.}
    \label{fig:bi-tri-comp}
\end{figure}

We now consider the possible role of inter-layer interactions by looking at the electronic structure of slabs with different number of TaSe$_3$ layers. 
We remind that the hump-shaped band near the Fermi level, which plays an important role in the competition between metallic and semiconducting states, has also an important component of the outer Se 4$p_y$ orbitals which are perpendicular to the TaSe$_3$ layers and thus, may alter the shape of this band. 
A comparison between Figs.~\ref{fig:mono-bulk-comp}a and b shows that both along the B $\rightarrow$ A and $\Gamma \rightarrow$ Y directions, the maximum of the band is kept below the Fermi level: in the region around $\Gamma \rightarrow$ Y is kept slightly below it, whereas around the B $\rightarrow$ A region is considerably lowered. This suggest that increasing the number of layers should not noticeably alter the behaviour analyzed for the case of a single-layer. However, to check that this is also the case for a small number of layers we have explicitly considered several slabs with different layers. Shown in Fig.~\ref{fig:bi-tri-comp} are the band structures for the optimized structures of slabs with two (a) and three (b) layers showing that even for a very few number of layers the essential details of the band structure, with the top of the hump-shaped band being a bit higher in the region around $\Gamma \rightarrow$ Y and staying just below the Fermi level, are kept. Comparing the band structures of Figs.~\ref{fig:mono-bulk-comp}b,~\ref{fig:bi-tri-comp}a and~\ref{fig:bi-tri-comp}b one can appreciate a slight increase of the semimetallic overlap but no major differences in the band overlap. Looking at the [Se$_I$ 4$p_y$]$_{pair}$ and Ta$_{II}$ 5$d_{z^2}$ panels of Fig.~\ref{fig:fatbands_reduced} it is clear that in the region around the top of the hump-shaped band in the $\Gamma \rightarrow$ Y direction there is a noticeable hybridization between the two type of orbitals. It turns out that the shorter Se$\cdots$Se interlayer contacts occur for one Se atom of chain I in one layer and one of chain II in the next layer which are just in front of each other. These contacts (one per each pair of chains) are shorter than twice the van der Waals radius of Se (for instance in the experimental bulk structure of TaSe$_3$ they are 3.60 \AA~and 3.64 \AA) and consequently, provide an additional coupling between chains I and II, i.e. an interlayer coupling. However, since the crossing is avoided in most of the Brillouin zone and the contacts are still relatively long, the top of the hump-shaped band never crosses the Fermi level. We have also verified that introduction of spin-orbit coupling in the calculations for bilayers and trilayers does not modify the results (Fig. S8 in SI). 

\section{Conclusions}

The electronic properties of transition metal trichalcogenides have brought considerable attention recently.
In this work we find that, among the handful of materials in this family, single-layer (or few-layer) TaSe$_3$ could be a versatile system amenable to interesting electronic variations. 
The metallic state of the pristine single-layer is modified in different ways under application of biaxial strain or uniaxial strain along either the short or long directions of the layer. 
Electronic instabilities related to both electron and hole pockets, as well as stabilization of a semiconducting state, are possible. 
Furthermore, the semiconducting gap can be tuned by changing the level of tensile strain along the long direction of the layer. 
The fact that bulk TaSe$_3$ exhibits superconductivity makes the experimental study of single- and few-layers TaSe$_3$ even more challenging.  
In view of the present results theoretical studies of TaSe$_3$ nanoribbons and microwires could also be very interesting.

\section{Computational details}\label{sec:methods}

First-principles calculations were carried out using a numerical atomic orbitals approach to DFT~\cite{kohsha1965,HohKoh1964}, which was developed for efficient calculations in large systems and implemented in the \textsc{Siesta} code~\cite{ArtAng2008,SolArt2002}. 
We have used the generalized gradient approximation (GGA) and, in particular, the functional of Perdew, Burke and Ernzerhof~\cite{PBE96}. 
Only the valence electrons are considered in the calculation, with the core being replaced by norm-conserving scalar relativistic pseudopotentials~\cite{tro91} factorized in the Kleinman-Bylander form~\cite{klby82}. 
The non-linear core-valence exchange-correlation scheme~\cite{LFC82} was used for all elements. 
We have used a split-valence double-$\zeta $ basis set including polarization functions~\cite{arsan99}. The energy cutoff of the real space integration mesh was set to 500 Ry. 
To build the charge density (and, from this, obtain the DFT total energy and atomic forces), the Brillouin zone (BZ) was sampled with the Monkhorst-Pack scheme~\cite{MonPac76} using grids of (14$\times$56$\times$1) and (14$\times$56$\times$14) {\it k}-points for the monolayer and bulk calculations, respectively. 
For the Fermi surface calculation a finer grid of  (28$\times$112$\times$1) and (28$\times$112$\times$28) {\it k}-points was used for the monolayer and bulk, respectively and were plotted using~\cite{xcrysden}. The structures were fully optimized (and visualized with~\cite{vmd}.
The biaxial strain was applied in both the $a$ and $b$ directions. 
The atomic positions were relaxed after applying the strains to the single-layer. 
The strain is defined as $s = \delta m/m_0$ where $m_0$ is the unstrained cell parameter and $\delta m + m_0$ the strained cell parameter. 
Thus, positive values correspond to tensile strain whereas negative numbers correspond to compressive strain.

\section*{Acknowledgments}
Numerical calculations presented in this paper have been performed on a supercomputing system in the Supercomputing Center of Wuhan University.
Work at Bellaterra (Spain) was supported by the Spanish MICIU through Grant PGC2018-096955-B-C44 and the MINECO through the Severo Ochoa Centers of Excellence Program under Grant SEV-2015-0496 as well as by Generalitat de Catalunya (2017SGR1506).

\bibliographystyle{apsrev4-1}
\bibliography{tase3_bib} 

\end{document}